\documentclass{aa}  
\bibpunct{(}{)}{;}{a}{}{,}

\usepackage{graphicx}
\usepackage{txfonts}
\usepackage{enumitem}
\usepackage{adjustbox}
\usepackage{url}
\usepackage[colorlinks=True,allcolors=blue,]{hyperref}
\usepackage{siunitx}
\begin{document} 
   \title{The X-ray/UV ratio in Active Galactic Nuclei: dispersion and variability \thanks{Table 2 is only available in electronic form at the CDS via anonymous ftp to \url{cdsarc.u-strasbg.fr} (130.79.128.5)}}

   \subtitle{}

   \author{E. Chiaraluce
          \inst{1,2},
          F. Vagnetti\inst{2},
          F. Tombesi\inst{2,3,4},
          \and
          M. Paolillo\inst{5,6,7}
          }

   \institute{INAF - Istituto di Astrofisica e Planetologia Spaziali (IAPS-INAF), Via del Fosso del Cavaliere 100, 00133 Roma, Italy\\
              \email{elia.chiaraluce@iaps.inaf.it}
         \and
             Dipartimento di Fisica, Università di Roma “Tor Vergata”, Via della Ricerca Scientifica 1, I-00133, Roma, Italy
         \and 
             X-ray Astrophysics Laboratory, NASA Goddard Space Flight Center, Greenbelt, MD 20771, USA
         \and 
             Department of Astronomy, University of Maryland, College Park, MD 20742, USA
          \and 
             Dipartimento di Fisica “Ettore Pancini”, Università di Napoli Federico II, via Cintia, I-80126 Napoli, Italy
          \and 
             INFN – Unita` di Napoli, via Cintia 9, I-80126 Napoli, Italy
          \and 
             Agenzia Spaziale Italiana – Science Data Center, Via del Politecnico snc, I-00133 Roma, Italy   
             \             }

   \date{Received ; accepted }

 
  \abstract
   {The well established negative correlation between the $\alpha_{OX}$ spectral slope and the optical/UV luminosity, a by product of the relation between X-rays and optical/UV luminosity, is affected by a relatively large dispersion. The main contributions can be variability in the X-ray/UV ratio and/or changes in fundamental physical parameters.}
   {We want to quantify the contribution of variability within single sources (\textit{intra-source} dispersion) and that due to variations of other quantities different from source to source (\textit{inter-source} dispersion).}
   {We use archival data from the XMM-Newton Serendipitous Source Catalog (XMMSSC) and from the XMM-OM Serendipitous Ultra-violet Source Survey (XMMOM-SUSS3). We select a sub-sample in order to decrease the dispersion of the relation due to the presence of Radio-Loud and Broad Absorption Line objects, and to absorptions in both X-ray and optical/UV bands.
   We use the Structure Function (SF) to estimate the contribution of variability to the dispersion. We analyse the dependence of the residuals of the relation on various physical parameters in order to characterise the inter-source dispersion.}
   {We find a total dispersion of $\sigma\sim$0.12 and we find that intrinsic variability contributes for 56$\%$ of the variance of the $\alpha_{OX} - L_{UV}$ relation. If we select only sources with a larger number of observational epochs ($\ge\,$3) the dispersion of the relation decreases by approximately 15\%. 
   We find weak but significant dependences of the residuals of the relation on black-hole mass and on Eddington ratio, which are also confirmed by a multivariate regression analysis of $\alpha_{OX}$ as a function of UV luminosity and black-hole mass and/or Eddington ratio.
   We find a weak positive correlation of both the $\alpha_{OX}$ index and the residuals of the $\alpha_{OX} - L_{UV}$ relation with inclination indicators, such as the FWHM(H$\beta$) and the EW[O$_{III}$], suggesting a weak increase of X-ray/UV ratio with the viewing angle. This suggests the development of new viewing angle indicators possibly applicable at higher redshifts. Moreover, our results suggest the possibility of selecting a sample of objects, based on their viewing angle and/or black-hole mass and Eddington ratio, for which the $\alpha_{OX} - L_{UV}$ relation is as tight as possible, in light of the use of the optical/UV - X-ray luminosity relation to build a distance modulus (DM) - $z$ plane and estimate cosmological parameters.  
   }
   {}

   \authorrunning{E. Chiaraluce et al.}
   \titlerunning{X-ray/UV ratio AGN}

   \maketitle
%

\section{Introduction}

The X-ray/UV ratio is a powerful tool which can be used to investigate distribution of Active Galactic Nuclei (AGN) X-ray and optical/UV properties \citep{LussoRisaliti2016,AvniTananbaum,Strateva2005} and their dependence on fundamental quantities like Eddington ratio, black-hole mass and redshift. The X-ray/UV ratio is usually defined in terms of the $\alpha_{OX}$ index as
\begin{equation}
\alpha_{OX}=\log{\frac{L(\nu_X)}{L(\nu_{UV})}}/\log{\frac{\nu_X}{\nu_{UV}}}
\end{equation}
but it is not rare to find it defined with a minus sign \citep[e.g.][]{Tananbaum1979,LussoRisaliti2016}, 
and it is usually considered $2\,$keV for the X-ray frequency and $2500{\si{\angstrom}}$ for the Optical/UV frequency \citep[e.g.][]{Vagnetti2010,LussoRisaliti2016}.
The $\alpha_{OX}$ index can be thought as the energy index or slope associated to a power law connecting the X-ray and Optical/UV bands \citep{Tananbaum1979}. 

In literature it has been studied the dependence of the X-ray/UV ratio on redshift, finding no significant dependence \citep[e.g.][]{VignaliBrandtSchneider2003, Strateva2005, Steffen2006, Just2007, Vagnetti2010, Vagnetti2013, DongGreenHo32012}. This means that energy generation mechanisms have not changed from early epochs: already at high redshift, AGNs were almost completely built-up systems, notwithstanding short available time to grow (Vignali, Brandt \& Schneider 2003; Strateva et al. 2005; Just et al. 2007). This picture is consistent with studies finding no significant evolution in AGNs continuum shape even at high redshift from radio \citep{Petric2003}, Optical/UV \citep{Pentericci2003} and X-ray \citep{Page2005}. 

The $\alpha_{OX}$ dependence on other parameters is still matter of debate. Some authors have found a significant correlation with the Eddington ratio $L/L_{Edd}$ \citep{Lusso2010} while other authors find no significant correlation with $L/L_{Edd}$ \citep{DongGreenHo32012, Vasudevan2009} and a significant one with $M_{BH}$ \citep{DongGreenHo32012}. 

It has been found in literature a strong, non-linear correlation between the X-ray/UV ratio and the monochromatic UV luminosity at $2500{\si{\angstrom}}$ in the form $\alpha_{OX}=a{\log{L_{UV}}}+b$, with $a$ in the interval $\sim\,-0.2\div-0.1$. However, this anti-correlation is the by-product of the well-established positive non-linear correlation between X-ray and Optical/UV luminosity ${L_X}{\propto}L_{UV}^{\gamma}$ with ${\gamma}\sim{0.5}\div{0.7}$ \citep[e.g.][]{AvniTananbaum, VignaliBrandtSchneider2003, Strateva2005, Steffen2006, Just2007, GibsonBrandtSchneider2008, Lusso2010, Vagnetti2010, Vagnetti2013, LussoRisaliti2016}.
Moreover, \citet{Buisson} analysed the variable part of the
UV and X-ray emissions for a sample of 21 AGN, finding that they are
also correlated with slopes similar to those found for the average
luminosities.

These two relations are symptoms of a tight physical coupling between the two regions responsible for the Optical/UV and X-rays, i.e. the accretion disk and X-ray corona, respectively. 
Indeed, standard accretion disk-corona models postulates an interaction between photons emitted from the accretion-disk and a central plasma of relativistic electrons constituting the corona, responsible for the emission of X-rays radiation.
Following standard picture by Haardt \& Maraschi (1991,1993), the soft thermal photons from disk, parametrised by $L_{2500{\si{\angstrom}}}$, are energised to X-rays by means of inverse Compton scattering on hot ($T_e\sim10^8\,K$) corona electrons, resulting in a power-law like component observed in AGNs X-ray spectra, with a cut-off corresponding to electron temperature \citep[e.g.][]{LussoRisaliti2016,Tortosa2018}. In this picture, the study of the $\alpha_{OX} - L_{UV}$ relation, or equivalently of the ${L_X} - {L_{UV}}$ relation, is of fundamental importance as we still lack a quantitative physical model explaining the existence of this correlation. However, in a recent paper \citet{LussoRisaliti2017} advanced a simple, \textit{ad-hoc} physical model for the accretion disk-corona system, predicting a dependence of the X-ray monochromatic luminosity on the monochromatic UV luminosity and the emission line full-width at half maximum of the form $L_X{\propto}{{L_{UV}}^{4/7}}{{v_{FWHM}}^{4/7}}$. Their model is based on accretion disk-corona models by \citet{SvZd1994}, in which magnetic loops and reconnection events above a standard Shakura-Sunyaev \citep{ShakuraSunyaev1973} accretion disk may be responsible for the emission of X-ray radiation \citep{LussoRisaliti2016}.

The $\alpha_{OX} - L_{UV}$ and $L_{X} - L_{UV}$ relations are however characterised by dispersion due to several causes: the Radio-Loud (RL) and Broad Absorption Lines (BAL) nature of some AGN, host galaxy effects, variability \citep{LussoRisaliti2016} (see Section \ref{sec:dispersion} for an extended discussion). AGN are variable in both Optical/UV and X-rays band. In the Optical/UV range many authors have confirmed variability \citep[e.g.][]{Cristiani1996,Giallongo1991,diClemente1996}, and the most reliable hypothesis is that of accretion disk instabilities \citep[e.g.][]{Vandenberk2004}. Variability in the X-rays band has been extensively studied with different methods like fractional variability \citep{Almaini, Manners2002}, the Power Spectral Density \citep{Papadakis2004, ONeill2005, UttleyMcHardy2005, McHardy2006,Paolillo2018}, the SF \citep{Vagnetti2011, Vagnetti2016, Middei2017}, and these works indicate that variations occur preferentially at long timescales  \citep[e.g.][]{Middei2017}. Variability is a major source of scatter in the above relations, and, once simultaneous observations are selected, its contributions reduces to essentially two factors: an intrinsic variations in the X-ray/UV ratio for single sources, and inter-sources variations. Previous works have estimated the contribution of the intrinsic variability in X-ray/UV ratio to the total variance of $\alpha_{OX} - L_{UV}$ relation to be roughly $\sim{30}\div{40\%}$ \citep{Vagnetti2010,Vagnetti2013}, but we still lack a physical explanation for the residual dispersion, and in this work we want to spread light on this topic.

In recent period, the study of the $L_X - L_{UV}$ relation has become more and more important as it has been used to built up a Hubble diagram for Quasars \citep{Risaliti2015,BisogniRisalitiLusso2017}. In order to achieve such a goal, the dispersion of the relation must be reduced as much as possible, and \citet{LussoRisaliti2016} proved that it is possible to do that by carefully selecting the sample. The use of this relation represents a valid alternative to the supernovae, as it can be used at higher redshift and it has a better statistics, but it has also shortcomings, as it relies on the tightness of the relation. For this very reason, a thorough study of the relation and of the physical origin of its dispersion is of fundamental importance, as it will aid in the selection of a sample of objects suited for the construction of a Hubble diagram.

In section 2 and 3 we describe the data from which we derived the sample we work with, in section 4 and 5  we describe the data analysis procedure together with results, in section 6 we discuss implications of our results in light of present and past works in literature. 

Throughout the paper we use a $\Lambda$-CDM cosmological model: $H_{0}=70\,km\,{s^{-1}}{Mpc^{-1}}$, $\Omega_m=0.3$ and $\Omega_{\Lambda}=0.7$.


\section{The Data}
The X-ray data used in this work come from the \textit{Multi-epoch X-ray Serendipitous AGN Sample} (MEXSAS2) catalogue \citep[][Vagnetti et al in preparation]{Serafinelli2017Frontiers}. The MEXSAS2 is a catalogue of 9735 XMM-\textit{Newton} observations for 3366 unique sources derived from the DR6 of the XMM-\textit{Newton} Serendipitous Source Catalogue \citep{Rosen2016} which have been identified with AGNs from SDSS DR7Q \citep{Schneider2010} and SDSS DR12Q \citep{Paris2017} quasar catalogues; it is an update of the MEXSAS catalogue defined in \citet{Vagnetti2016}. The MEXSAS2 catalogue provides black-hole mass, Eddington ratio and bolometric luminosity by cross-match it with two catalogues of quasar properties published by \citet{Shen2011} and \citet{Kozlowski2017}. We caution that the black-hole mass estimates are to be considered with a typical uncertainty of 0.4 dex, and the bolometric luminosities have been derived from bolometric corrections which are only appropriate in a statistical sense, as discussed by \citet{Shen2011}.

In order to perform an X-ray/UV ratio variability study, the MEXSAS2 catalogue has been cross-matched with the XMM-SUSS3, the third version of the XMM-OM Serendipitous Ultraviolet Source Survey \citep{Page2012}, based on the XMM-\textit{Newton} satellite there is the Optical Monitor, an Optical/UV telescope with a primary mirror of 30 cm \citep{Mason2001}. The XMMOM-SUSS3 provides fluxes in six filters, i.e. UVW2, UVM2, UVW1, U, B, V, with central wavelengths $1894{\si{\angstrom}}$, $2205{\si{\angstrom}}$, $2675{\si{\angstrom}}$, $3275{\si{\angstrom}}$, $4050{\si{\angstrom}}$ and $5235{\si{\angstrom}}$, respectively (see the dedicated page at MSSL\footnote{\url{http://www.mssl.ucl.ac.uk/~mds/XMM-OM-SUSS/SourcePropertiesFilters.shtml}}).
In the XMM-SUSS3, many sources are observed more than once per filter, and this allows to perform variability studies.

The cross-match between the MEXSAS2 catalogue and the XMMOM-SUSS3 has been performed using 1.5 arcsec as matching radius and then comparing the OBS{\textunderscore}ID and OBSID flags in 3XMM-DR6 and XMMOM-SUSS3 with the Virtual Observatory software TOPCAT \footnote{\url{http://www.star.bris.ac.uk/~mbt/topcat/}} \citep{Topcat}: in this way we impose that matched X-ray and UV entries from XMM-\textit{Newton} and XMMOM-SUSS3 catalogues correspond to the same observation.

The result of the cross match consists of 1857 observations for 944 unique sources, 438 of which are single-epoch, the remaining ones are multi-epoch. Note that, although e started with a multi-epoch catalogue, MEXSAS2, after the cross-match with XMMOM-SUSS3 we ended up with a sample of both single-epoch and multi-epoch sources. This is due to the cross-matching procedure. Indeed, the Optical Monitor for the Optical/UV measurements is co-axial with the Epic Cameras for the X-ray measurements, but the two instruments have different FoVs: $17\,arcmin^2$ and $30\,arcmin^2$, respectively. This explains the reduced number of observations and the presence of single epoch sources in the sample.
The XMM-SUSS has been available from the XMM-SUSS page \footnote{\url{http://www.ucl.ac.uk/mssl/astro/space_missions/xmm-newton/xmm-suss3}}, the XMM-\textit{Newton} Science Archive \footnote{\url{https://www.cosmos.esa.int/web/xmm-newton/xsa}} and the NASA High Energy Astrophysics Science Archive Research Center HEASARC \footnote{\url{http://heasarc.gsfc.nasa.gov}}.

\subsection{UV and X-ray luminosities}

In order to calculate the X-ray/UV ratio we need to determine the X-ray and UV rest-frame luminosities. It is customary to choose the $2\,keV$ and $2500{\si{\angstrom}}$ luminosities as representatives of the two quantities.

Considering UV measurements, for each object we can have one or more estimates of fluxes from one up to six OM filters. Considering a single object and a single observation, we can calculate the rest-frame monochromatic UV luminosity corresponding to each of the OM filters:
\begin{equation}
L_{\nu_{}}(\nu_{em})=F_{\nu_{}}(\nu_{obs})\frac{4{\pi}{{D_L}^2}}{1+z}
\end{equation}
where $D_L$ is the luminosity distance of the source at redshift $z$, $F_{\nu}(\nu_{obs})$ is the observed flux in one of the six OM filters. In this way it is possible to build individual Spectral Energy Distributions (SEDs) in the UV for the objects in the sample. 

\begin{figure}
\includegraphics[width=0.49\textwidth]{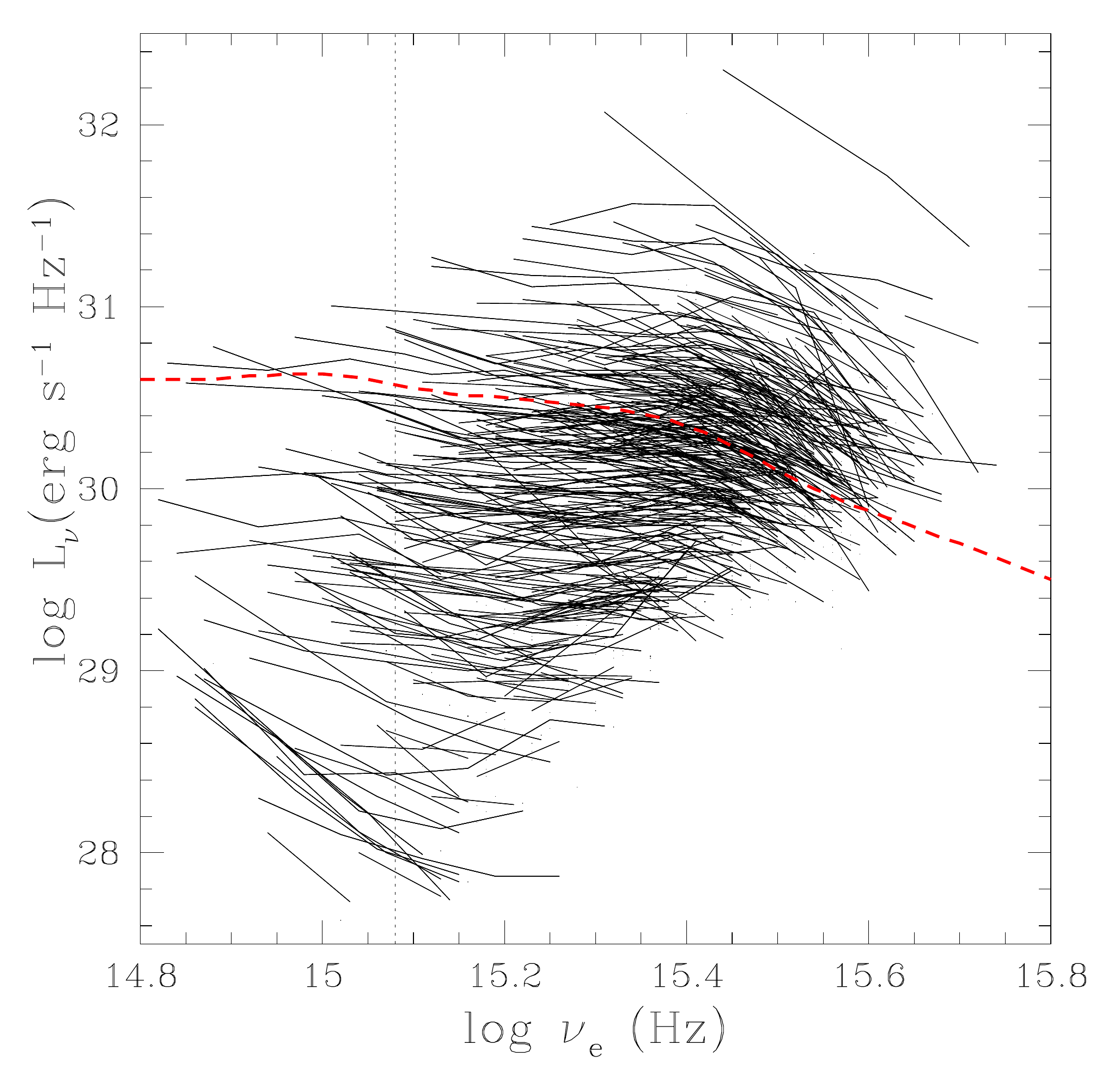}
\caption{SEDs averaged in time for the multi-epoch objects in the parent sample: for each frequency we consider the average of the quantities $L_{\nu_{}}(\nu_{em})$. The red dashed line is the average SED by \citet{Richards2006a} for type-1 objects in the SDSS. The vertical line is at $\log{\nu_{e}}=15.08$, corresponding to $2500\,\si{\angstrom}$.}
\label{fig:sed_average}
\end{figure}

In Figure \ref{fig:sed_average} we show average SEDs: for each frequency we consider the average $L_{\nu_{}}(\nu_{em})$ over all observations.

The rest frame monochromatic UV luminosity $L_{2500{\si{\angstrom}}}$ is derived with the procedure adopted by \citet{Vagnetti2010}, which can be summarised in the following way: i) in the case in which the available $L_{\nu}(\nu_{em})$ estimates from the OM filters cover only frequencies higher or lower than $2500{\si{\angstrom}}$ ($\log{\nu_{2500{\si{\angstrom}}}}=15.08$, the vertical dashed line in Figure \ref{fig:sed_average}) the $L_{2500{\si{\angstrom}}}$ is calculated through curvilinear extrapolation, following the behaviour of the average UV SED by \citet{Richards2006a}, computed for type-1 objects in the SDSS, shifted vertically to match the luminosity of the frequency of the nearest point; ii) if the SED extends across $\log{\nu_{2500{\si{\angstrom}}}}=15.08$, the $L_{2500{\si{\angstrom}}}$ is calculated as linear interpolation of two nearest SED points; iii) if $L_{\nu}(\nu_{obs})$ is measured at only one frequency, $L_{2500{\si{\angstrom}}}$ is calculated as in (i). We note that there are some cases of anomalous and steep SEDs at high luminosities and frequency, possibly affected by intergalactic HI absorption, which will be removed according to the discussion in Section 3.1, and at low luminosities and frequencies, where the contribution of the host galaxy can be important. In both cases, we assume that the intrinsic SED is similar to the average SED of quasars according to \citet{Richards2006a} and our extrapolation is performed from a frequency which is relatively close to 2500 ${\si{\angstrom}}$.

The UV data has been then corrected for extinction following \citet{LussoRisaliti2016}. The galactic extinction is estimated from \citet{Schlegel1998} for each object \footnote{\url{http://irsa.ipac.caltech.edu/applications/DUST/}} while the normalised selective extinction has been estimated for each filter as linear interpolation of mean extinction curve by \citet{Prevot1984}.

\begin{figure}
\includegraphics[width=0.5\textwidth]{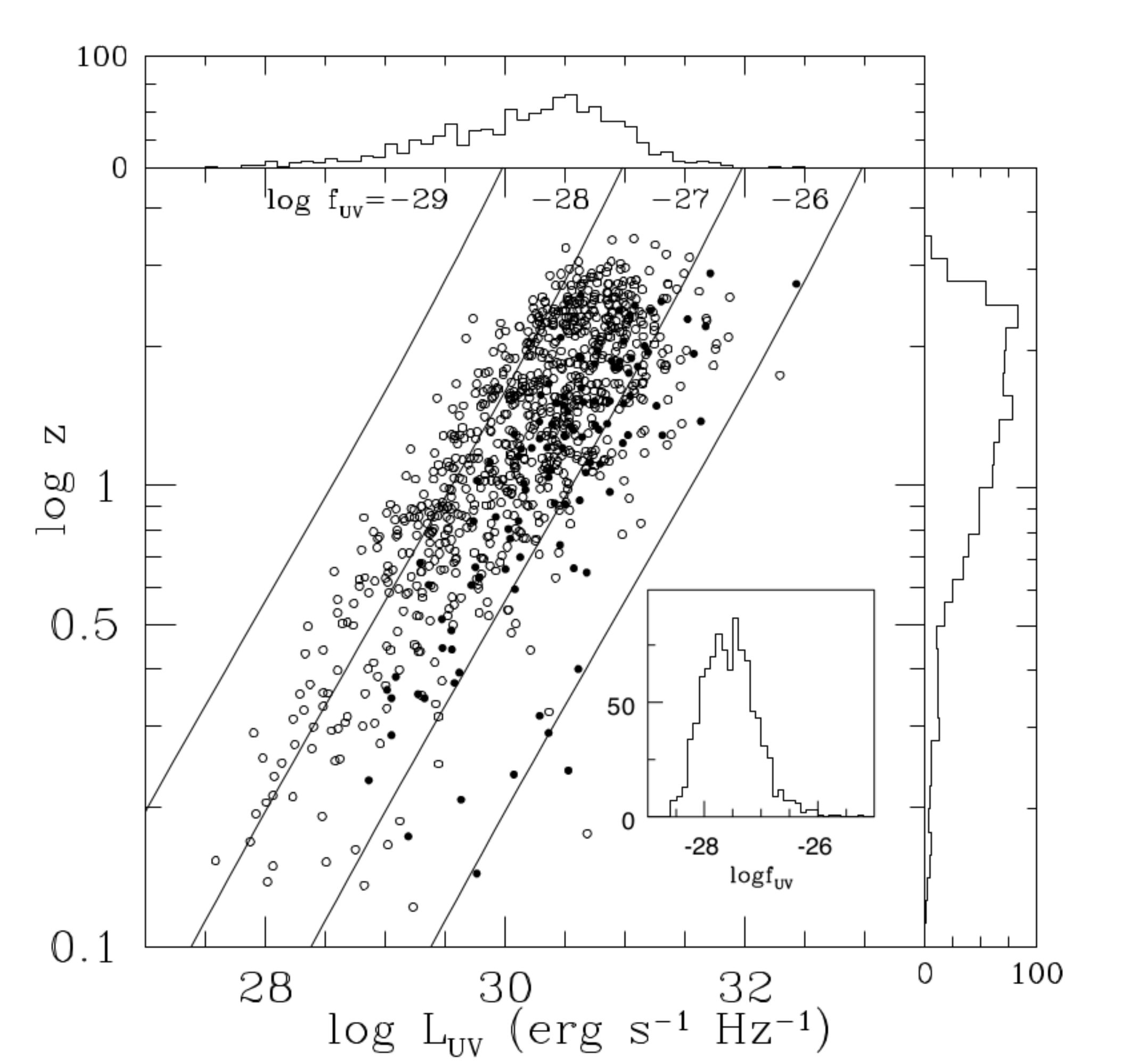}
\caption{Distribution of the parent sample in the $L_{UV}$-$z$ plane. Straight lines are lines of constant UV flux. Black circles represent objects in common with \citet{Vagnetti2010}, empty circles are objects present only in this work. The straight lines are constant UV flux lines. The histogram on top shows the distribution of objects in the parent sample with respect to UV luminosity; the histogram on the right shows the redshift distribution of objects in the parent sample. The histogram inside the plot area (lower-right) shows the distribution of objects in the parent sample with respect to UV flux.}
\label{fig:pianoLUVz}
\end{figure}

The sample described so far is referred to as 'parent' sample, and in Figure \ref{fig:pianoLUVz} it is shown the distribution of the parent sample in the $z$-$L_{UV}$ plane.

To calculate the rest-frame monochromatic $2\,keV$ luminosity, we started from fluxes in the XMM-SSC DR6 energy bands: EP{\textunderscore}2, EP{\textunderscore}3, EP{\textunderscore}4 and EP{\textunderscore}5, in the intervals $0.5-1.0\,keV$; $1.0-2.0\,keV$; $2.0-4.5\,keV$ and $4.5-12.0\,keV$, respectively. We calculated the $2\,keV$ luminosity performing the procedure adopted by \citet{LussoRisaliti2016}. We combined band 2 and 3 to form a 'soft' band by simply summing fluxes in two bands, with uncertainty summed in quadrature, the same has been done to form a 'hard' band from band 4 and band 5. The resulting bands are therefore in the intervals $0.5-2.0\,keV$ (Soft Band) and $2.0-12.0\,keV$ (Hard Band):

In each band, we have first assumed a power-law with a typical photon index $\Gamma_0=1.7$, and then we have calculated the rest frame monochromatic luminosity at the frequency corresponding to the geometric mean of the band: $L_{\nu}(1\,keV)$ and $L_{\nu}(5\,keV)$ for the Soft and Hard bands, respectively. 

Once $L_{\nu}(1\,keV)$ and $L_{\nu}(5\,keV)$ have been calculated, they have been used to derive an estimate of the photon index $\Gamma$ by assuming a power-law connecting the two bands:
\begin{equation}
{1-{\Gamma}}=\frac{\log{L_{\nu}^{}(5\,keV)}-\log{L_{\nu}^{}(1\,keV)}}{\log{({\nu_{5\,keV}}/{\nu_{1\,keV}})}}
\label{eqn:GammaC}
\end{equation}
This $\Gamma$ has been then used to determine the rest-frame monochromatic $2\,keV$ luminosity.

\section{The $\alpha_{OX}-{L_{UV}}$ relation and its dispersion}
\label{sec:dispersion}
The $\alpha_{OX} - L_{UV}$ relation, and the $L_{X} - L_{UV}$ relation, are characterised by relatively large dispersions, of roughly $0.13-0.15\,$dex \citep[e.g.][]{Strateva2005,Just2007,GibsonBrandtSchneider2008} and $0.35-0.4$dex \citep{LussoRisaliti2016} respectively. The main factors contributing to the dispersion concern the nature of the objects, the emission properties, the host galaxies effects and the use of simultaneous X-ray and UV data \citep{LussoRisaliti2016}. 

Indeed, radio-loud objects would lie far above the average relation, because of their enhanced X-ray emission associated with jets \citep{Worrall1987}, resulting in higher X-ray/UV ratios at fixed UV luminosity. It is important to exclude them as their X-ray emission is not only the nuclear component.

Broad-absorption-line (BAL) quasars would contribute in the opposite sense to the dispersion, as they have lower X-ray/UV ratio at fixed UV luminosity. Although they are believed to be characterised by the same underlying continua, absorption is believed to make them X-ray weak \citep{VignaliBrandtSchneider2003,Gallagher2001,Gallagher2002,Green2001}, and this property is not dependent on redshift \citep{BrandtLaorWills2000,BrandtGuainazzi2001,Vignali2001,Gallagher2002}. 

Intrinsic X-ray weakness can contribute to the dispersion, as there is evidence for a significant population of Soft-X-ray-weak (SFX) objects \citep{Laor1997,Yuan1998} which may be caused by absorption, unusual SEDs and/or Optical/X-ray variability \citep{BrandtLaorWills2000}. 

Host galaxy starlight effect can be taken into account \citep[e.g.][]{Lusso2010,Vagnetti2013,LussoRisaliti2016}. \citet{Vagnetti2013} follow the same approach of \citet{Lusso2010}. The optical spectrum is modelled as a combination of host galaxy + AGN contribution: $L_{\nu}=A[{f_A}F_R(\nu)+{f_G}{(\nu/\nu^{*})^{-3}}]$ where $F_R$ is the mean SED by \citet{Richards2006a}; $\nu^{*}$ is the frequency corresponding to 2500 $\si{\angstrom}$, and $f_{A,G}$ represent the fractional contribution of AGN and galaxy, respectively, at 2500 $\si{\angstrom}$. The normalising constant A is determined in a self-consistent way.

The slope of the $\alpha_{OX} - L_{UV}$ relation corrected for the host galaxy contribution should be \textit{steeper} than the uncorrected one \citep{Wilkes1994}, although this effect should be more important for samples with a relevant number of low-luminosity objects \citep{Vagnetti2013}.

Variability can be an important factor contributing to the $\alpha_{OX} - L_{UV}$ relation dispersion. Variability in the $\alpha_{OX}$ index can be an artificial variability, due to non-simultaneity of UV and X-ray data, or an intrinsic variability, due to a true variability in the X-ray/UV ratio. It is possible to eliminate artificial variability by using simultaneous UV and X-ray data \citep{Vagnetti2010,Vagnetti2013,LussoRisaliti2016}, in order to directly investigate true variability in X-ray/UV ratio. However \citet{Vagnetti2010}, using simultaneous data, have found that the dispersion of the relation is not significantly different from that derived by other authors \citep{Strateva2005,Just2007,GibsonBrandtSchneider2008} using non-simultaneous data, and this result has been confirmed by \citet{LussoRisaliti2016}. 

Although our UV and X-ray measurements are simultaneous, the emission processes in these two bands occur in different regions, so we should take into account also the propagation times. However the X-ray-UV lags are estimated within a few days \citep[e.g.][]{Marshall2008,Arevalo2009}, which will be neglected compared to the year-long timescales of our $\alpha_{ox}$ variations, see Section \ref{SF}.

The observed dispersion may be due to two factors: an \textit{intra-source} dispersion and an \textit{inter-source} one, the former due to intrinsic variation of the X-ray/UV ratio for individual sources, the latter due to differences in the X-ray/UV ratio among different sources.

Considering the variability, which accounts for the \textit{intra-source} dispersion, we can have two scenarios. The first one refers to variability occurring on short timescales, days$\,\div\,$weeks, because of variations in the X-ray flux which irradiates the part of the disk responsible for the Optical/UV emission (so X-ray driven variations). The second one refers to perturbations in the outer accretion disk, which propagate inwards modulating, on long-timescales (months$\,\div\,years$), the X-ray emission through variations in the Optical/UV photons field, so optically-driven variations \citep{Lyubarskii1997, Czerny2004, ArevaloUttley2006, Papadakis2008, McHardy2010, Vagnetti2010, Vagnetti2013}. \citet{Vagnetti2010} and \citet{Vagnetti2013} found an increasing SF($\alpha_{OX}$) as a function of the time-lag, with variations occurring preferentially at long timescales, suggesting optically-driven variations.

\subsection{The Reference sample}

\begin{figure}
\includegraphics[width=0.52\textwidth]{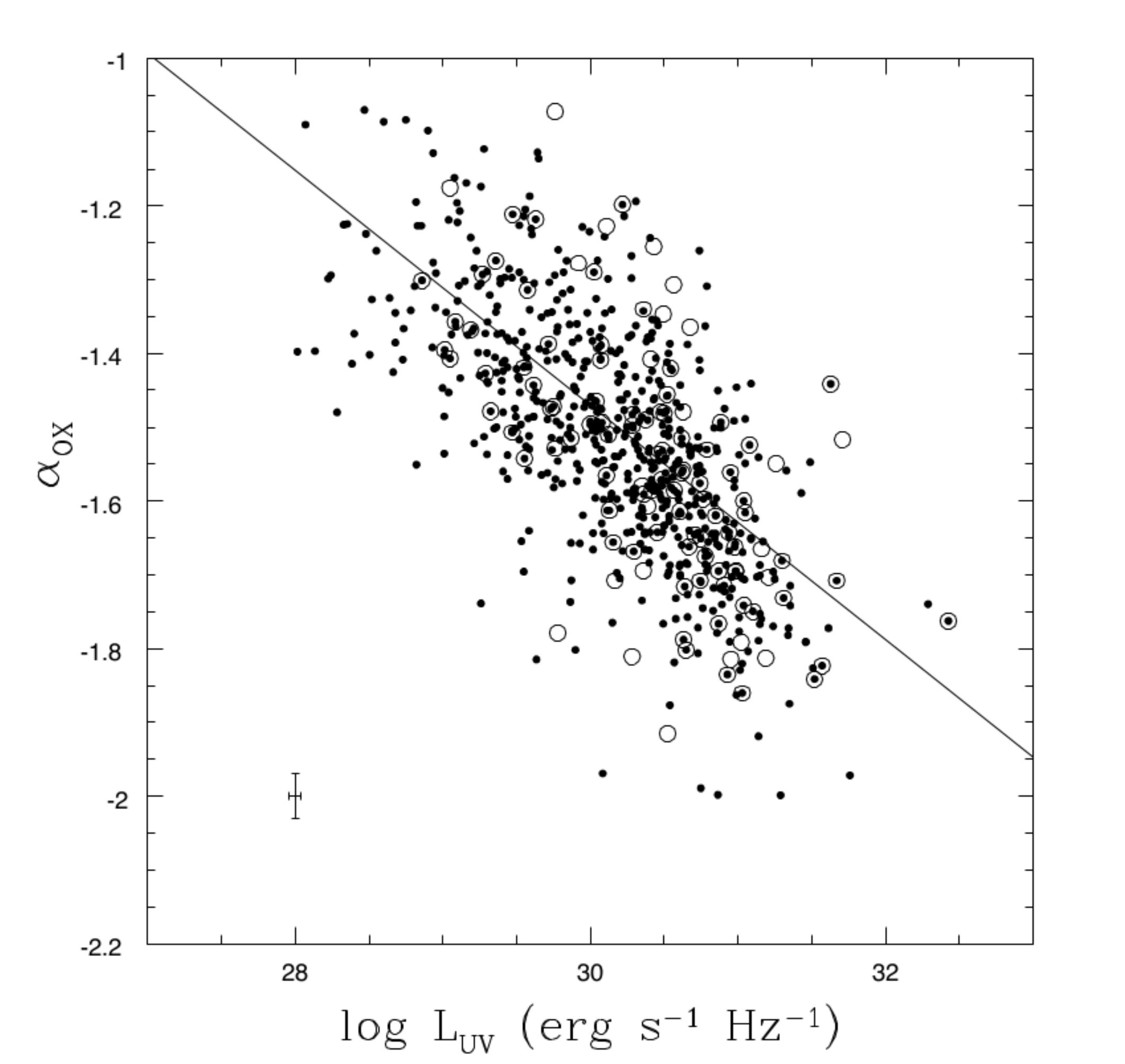}
\caption{$\alpha_{OX}$ as a function of the UV luminosity for the objects in the reference sample (black points) and for comparison empty circles represent objects from \citet{Vagnetti2010}; empty circles with a black point inside represent objects belonging to both groups. The straight line is the linear least squares fit to the data considering only the reference sample. Average uncertainties on the two quantities are shown with a representative point with error bars.}
\label{fig:graph_3}
\end{figure}

As outlined by \citet{LussoRisaliti2016}, it is possible to decrease the dispersion of the $L_X - L_{UV}$ relation, and in turn of the $\alpha_{OX} - L_{UV}$ relation, by carefully selecting the sample to work with. In \citet{LussoRisaliti2016}, this has been done in order to build a Hubble diagram for Quasars. Indeed, they use the $L_X - L_{UV}$ to build a $z -$DM diagram (DM being the distance modulus) for quasars, analogous to that of supernovae, to estimate cosmological parameters associated to $\Lambda$CDM cosmological models, but in order to get competitive results it is necessary to decrease as much as possible the dispersion of the $L_X - L_{UV}$ relation. We do not use here the $\alpha_{OX} - L_{UV}$ relation for cosmological applications, nevertheless we perform variability studies with a sample selected with same criteria used by \citet{LussoRisaliti2016} and compare our results with theirs.

As mentioned before, Radio-Loud and BAL sources would increase the dispersion of the $\alpha_{OX} - L_{UV}$ relation, lying far away from the average relation. In order to identify Radio-Loud sources, we computed the radio-loudness parameter \citep{Kellerman1989}:
\begin{equation}
R^{*}=\frac{L_\nu(5\,GHz)}{L_{2500\si\angstrom}}
\label{eqn:radioloudness}
\end{equation}
an object is identified as Radio-Loud if $R^{*}\,>\,10$, otherwise it is classified as Radio-Quiet. Indeed, objects  from SDSS-DR7 were already provided with the radio-loudness parameter, while for objects from SDSS-DR12 we calculated the radio flux density at $5\,GHz$ starting from radio flux density at $1.4\,GHz$ adopting a radio spectral index of $\alpha=-0.8$ {\citep[e.g.][]{GibsonBrandtSchneider2008}}. Both catalogues by \citet{Shen2011} and \citet{Kozlowski2017} flagged BAL sources, so we used their classification. However, the BAL nature is not always obvious, as BALs can appear and or disappear on month/year timescales, making them difficult to identify \citep{DeCicco2018}

We have also taken into account the intergalactic $H_{I}$ absorption, which would result in a suppression in the source flux at wavelengths smaller than the $Ly\alpha$ wavelength of $1216\,{\si{\angstrom}}$, and so in an underestimation in the UV luminosity. We essentially select only those objects whose SEDs are such that the nearest SED point to $\log{\nu_{em}}\,=\,15.08$ (corresponding to $2500\,{\si{\angstrom}}$) is at a frequency smaller than frequency corresponding to $1216\,{\si{\angstrom}}$: we exclude those objects for which the effect of intergalactic $H_{I}$ absorption should be significant. Then, we considered only non absorbed sources and only those ones having reasonable estimates of the photon index, with the conditions $1.6{\,}{\le}{\,}{\Gamma_C}{\,}{\le}{\,}2.8$ \& $\Gamma_C/\delta\Gamma_C>1.5$.

Therefore, the reference sample is defined by the set of conditions 
\begin{enumerate}[label=\roman{*}., ref=(\roman{*})] 
\item No Radio-Loud and No BAL sources
\item $\log{\nu_{em}^{nearest}}<15.4$
\item $1.6{\,}{\le}{\,}{\Gamma_C}{\,}{\le}{\,}2.8$ \& $\Gamma_C/\delta\Gamma_C>1.5$
\end{enumerate}
and it is constituted by 1095 observations corresponding to 636 sources, 273 of which are multi-epoch. In Table \ref{tab:tab_samples} we show the properties of both the Parent and the Reference sample.

The data of the Reference sample are reported in Table 2 (in electronic form \url{cdsarc.u-strasbg.fr}) where the columns are: \textit{Col.} (1), identification number of the source in the MEXSAS2 catalogue; \textit{Col.} (2), SDSS name; \textit{Cols} (3) \& (4), coordinates of the SDSS identification; \textit{Col.} (5), redshift; \textit{Col.} (6), black-hole mass; \textit{Col.} (7) bolometric luminosity; \textit{Col.} (8), Eddington ratio; \textit{Col.} (9), number of observations; \textit{Col.} (10), time of observation (MJD); \textit{Cols} (11) \& (12), log of the monochromatic luminosity at 2500 ${\si{\angstrom}}$ and its uncertainty; \textit{Cols} (13) \& (14), log of the monochromatic luminosity at 2 keV and its uncertainty; \textit{Cols} (15) \& (16), the $\alpha_{OX}$ index and its uncertainty.

\begin{center}
\centering
\begin{table}\small
\centering
\caption{{Summary of properties for both the Parent and Reference sample.}}
\begin{adjustbox}{width=0.5\textwidth}
\begin{tabular}{cccccc}
\hline
Sample & \# Observations & \# Sources & \# M.E. & \# M.E. (\# Obs $>$ 2) & \# S.E. \\ 
(1) & (2) & (3) & (4) & (5) & (6) \\ 
\hline
{Parent} &  1857 & 944 & 506 & 202 & 438 \\ 
{Reference} & 1095 & 636 & 273 & 92 & 363 \\
\hline
\end{tabular}
\end{adjustbox}
\tablefoot{{Column (1) Sample; column (2) Number of  observations; column (3) number of sources; column (4) number of multi-epoch (M.E.) sources; column (5) number of multi-epoch sources with a number of observations > 2; column (6) number of single-epoch (S.E.) sources}}
\label{tab:tab_samples}
\end{table}
\end{center}

\begin{figure}
\includegraphics[width=0.51\textwidth]{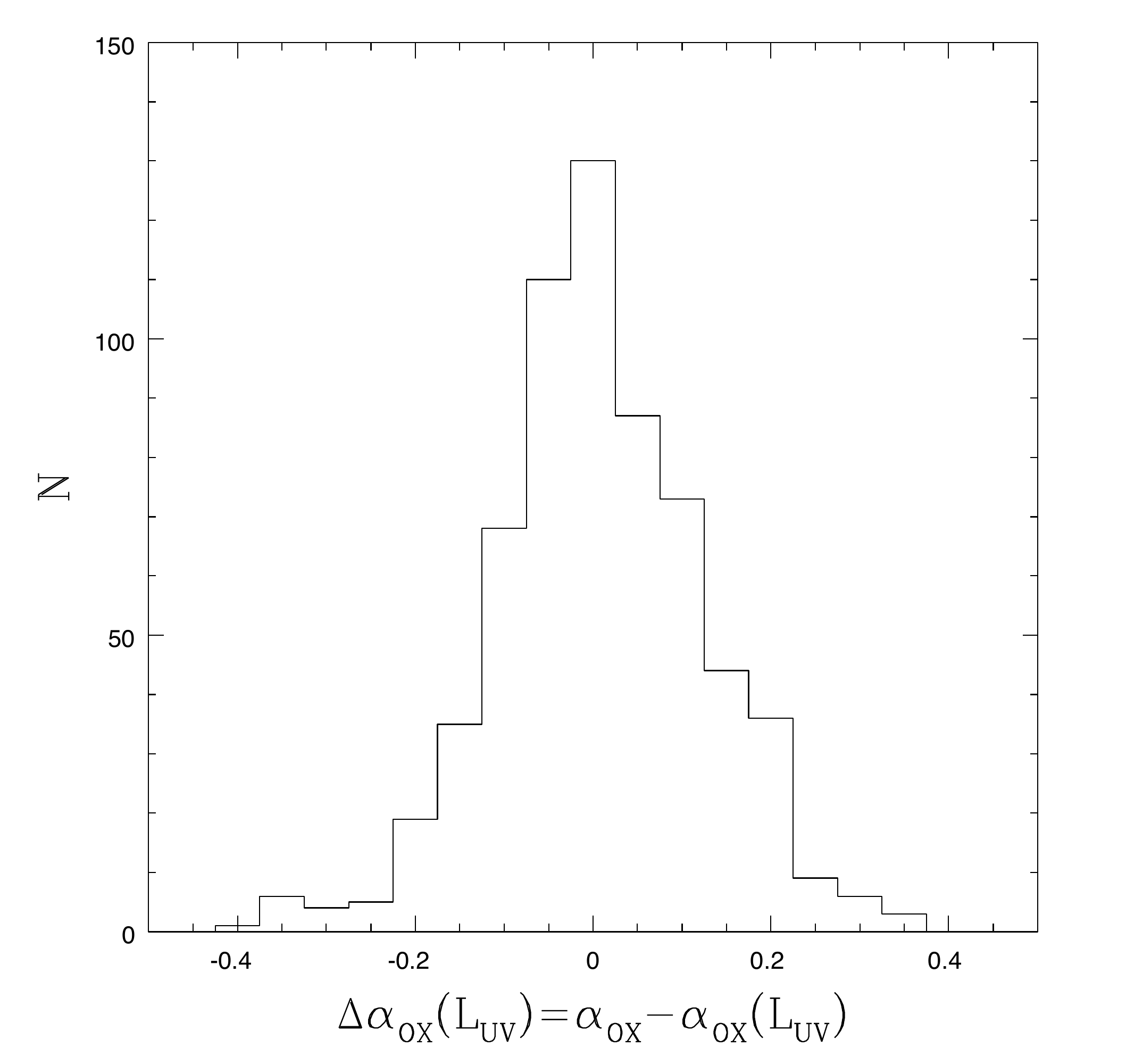}
\caption{Histogram showing the distribution of the residuals of the $\alpha_{OX} - L_{UV}$ relation for the objects in the reference sample, characterised by a standard deviation of ${\sigma=0.12}$.}
\label{fig:res_luv}
\end{figure}

With the sample described above we studied the $\alpha_{OX} - L_{UV}$ relation. In Figure \ref{fig:graph_3} we show the distribution of objects belonging to the reference sample in the $\alpha_{OX} - L_{UV}$ plane. Open circles are objects in common with \citet{Vagnetti2010}; black points represent objects belonging to the reference sample; open circles with black point inside represent objects in the reference sample which are in common with \citet{Vagnetti2010}. In figure \ref{fig:graph_3} it is also shown the linear least squares fit to the data for the 636 objects of the reference sample. In order to perform the linear fit, for a single-epoch object we considered the only available estimates of $\alpha_{OX}$ and $L_{UV}$, for a multi-epoch one we considered average values of the two quantities over the different epochs. The result of the fit is 
\begin{equation}
\alpha_{OX}=(-0.159{\pm}0.007)\log{L_{UV}}+(3.30{\pm}0.21)
\label{eqn:fit}
\end{equation}
with a correlation coefficient ${r=-0.69}$ and a probability ${P(>r)=2.7\times10^{-90}}$ for the null hypothesis that $\alpha_{OX}$ and $L_{UV}$ are uncorrelated.

In Figure \ref{fig:res_luv} it is shown the histogram of the distribution of the residuals of the of the $\alpha_{OX} - L_{UV}$ relation:
\begin{equation}
\Delta\alpha_{OX}=\alpha_{OX} - \alpha_{OX}(L_{UV})
\label{eqn:residuals}
\end{equation}
and it is characterised by a standard deviation ${\sigma=0.12}$.
 
As a comparison, we studied the $\alpha_{OX}-L_{UV}$ relation also for the parent sample, adopting the same procedure used for the reference sample, and we found $\alpha_{OX} =(-0.165\pm0.006)\log{L_{UV}} +(3.45\pm0.19)$, with a dispersion of $\sigma\sim$0.14. This means that the adopted strategy of selecting the sample according to the constraints described above actually translates into a decrease in the dispersion of the relation.

The value of $\sim$0.12 is consistent with previous works \citep{Strateva2005, Just2007, GibsonBrandtSchneider2008,Vagnetti2010}. Our slope, Equation \ref{eqn:fit}, is consistent with \citet{Vagnetti2010}, they obtained a correlation $\alpha_{OX}=(-0.166{\pm}0.012)\log{L_{UV}}+(3.489{\pm}0.377)$. Moreover, our slope can be compared with that obtained by previous works: it turns to be not consistent with \citet{GibsonBrandtSchneider2008}, who found $\alpha_{OX}=(-0.217{\pm}0.036)\log{L_{UV}}+(5.075{\pm}1.118)$, and with \citet{Grupe2010}, who found $\alpha_{OX}=(-0.114{\pm}0.014)\log{L_{UV}}+(1.177{\pm}0.305)$. However, as already pointed out by \citet{Vagnetti2010}, this may be due to the fact that they deal with samples of limited intervals of UV luminosity and/or redshift, and there is evidence of a dependence of the slope of the relation on these quantities (see detailed discussion in \ref{parametri}).

\begin{figure}
\includegraphics[width=0.5\textwidth]{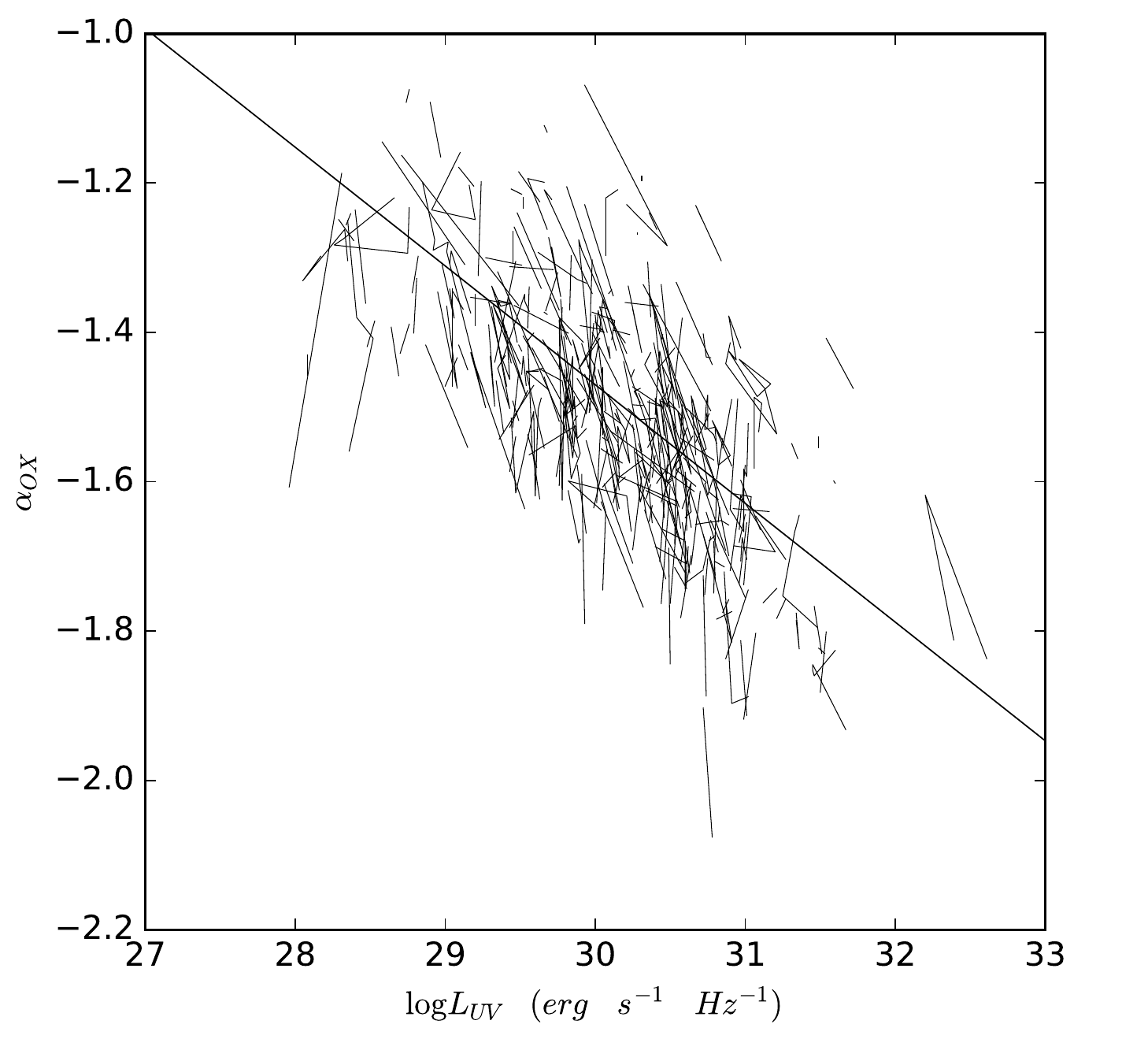}
\caption{$\alpha_{OX}$ index as a function of the UV luminosity for the objects in the reference sample in which tracks of multi-epoch objects in this plane are shown. The straight line is the linear least squares (equation \ref{eqn:fit}) to the data of the reference sample.}
\label{fig:tracks_campione_selezionato}
\end{figure}

We show in Figure \ref{fig:tracks_campione_selezionato} the tracks of individual objects of the reference sample in the $\alpha_{OX} - L_{UV}$ plane, clearly indicating the effect of variability on the dispersion of the observed relation.

In light of trends in Figure \ref{fig:tracks_campione_selezionato}, a possible way of reducing the dispersion of the relation would be to remove sources observed only few times (e.g. one or two epochs). Indeed, the estimate of the average values of $\alpha_{OX}$ (and also UV luminosity) are more robust considering a larger number of epochs. Excluding single-epoch objects, so considering only the 273 multi-epoch objects, we find $\alpha_{OX}=(-0.15\pm0.01)\log{L_{UV}}+(3.0\pm0.03)$, with r=0.71,  P($>$r)$\sim$2$\times$10$^{-43}$ and a dispersion $\sigma\sim$0.11. If we consider now 92 objects with three or more observations (see Table \ref{tab:tab_samples}), we find $\alpha_{OX}=(-0.14\pm0.02)\log{L_{UV}}+(2.7\pm0.5)$, with r=0.68, P($>$r)$\sim$1$\times$10$^{-13}$ and $\sigma\sim$0.10.

\subsection{Multi-epoch data: The Structure Function}
\label{SF}

The Structure Function has been extensively used in the literature to perform ensemble variability studies both in the Optical/UV band \citep[e.g.][]{Trevese1994, Cristiani1996, Wilhite2008, Bauer2009, MacLeod2012} and in the X-ray band \citep[e.g.][]{Vagnetti2011, Vagnetti2016, Middei2017} considering fluxes and magnitudes. The Structure Function (SF) gives a measure of variability as a function of time-lag $\tau$ between two observations. It can be used in principle to study the variability of any quantity, and it has been defined in different ways in literature \citep{Simonetti1985, diClemente1996}. In this work we adopt the definition by \citet{Simonetti1985}, which in the case of the $\alpha_{OX}$ rewrites as
\begin{equation}
SF(\tau)=\sqrt{{\langle}[{\alpha_{OX}}(t+\tau)-{\alpha_{OX}}(t)]^2{\rangle}-\sigma_n^2}
\label{eqn:sf}
\end{equation}
where $\sigma_n^2$ is the contribution of the photometric noise to the observed variability:
\begin{equation}
{\sigma_n^2={\langle}(\delta{\alpha_{OX}(t)})^2+(\delta{\alpha_{OX}(t+\tau)})^2{\rangle}{\sim}2{\langle}({\delta{\alpha_{OX}}})^2{\rangle}}
\end{equation}
with $\delta{\alpha_{OX}}$ being the uncertainty associated with $\alpha_{OX}$.
The plane is divided into bins of time-lag (in log units), and in each bin it is computed the ensemble average value of the square of the difference $\alpha_{OX}(t+\tau)-\alpha_{OX}(t)$, considering all the pairs of observations for each object lying in the relevant bin of time-lag $\tau$. The time-lag value representative of the bin is calculated weighting for the distribution of points within the bin.

The structure function can be used to put constraints on the contribution of variability to the total dispersion of the $\alpha_{OX} - L_{UV}$ relation. Indeed, following \citep{Vagnetti2010, Vagnetti2013}, it is possible to write the total variance of the $\alpha_{OX} - L_{UV}$ relation as the sum of two contributions (see section \ref{sec:dispersion}):
\begin{equation}
\sigma^2={{\sigma^2}_{intra-source}}+{{\sigma^2}_{inter-source}}
\end{equation}
From the SF value at long time-lags we can estimate the fractional contribution of the intra-source dispersion ${{\sigma^2}_{intra-source}}/\sigma^2$, i.e. the contribution of the true variation in the X-ray/UV ratio to the dispersion of the $\alpha_{OX} - L_{UV}$ relation. Previous works found it to be ${\sim}\,30\,{\%}$ \citep{Vagnetti2010} and ${\sim}\,40\,{\%}$ \citep{Vagnetti2013}.

\begin{figure}
\includegraphics[width=0.5\textwidth]{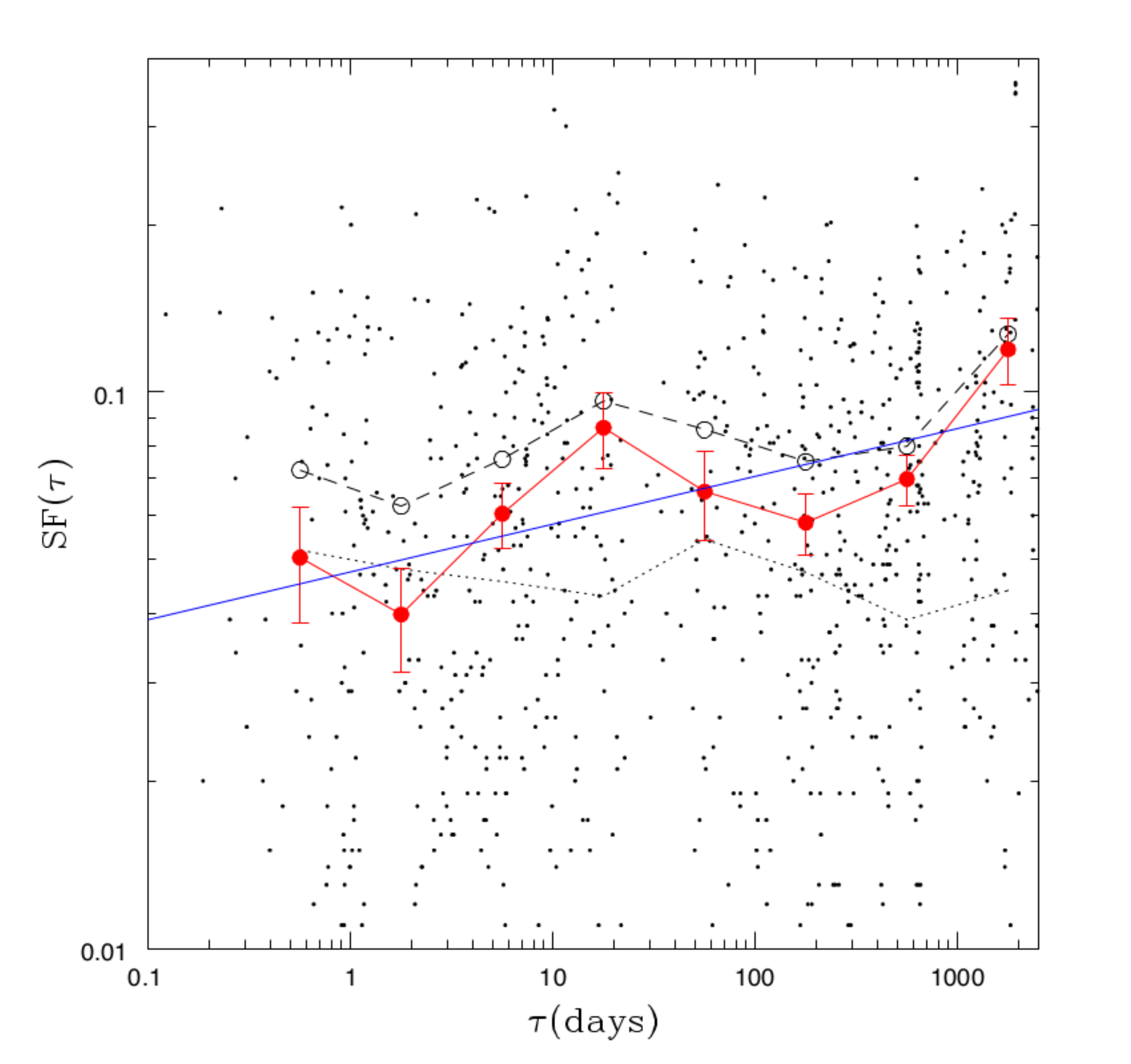}
\caption{Structure function ${SF(\tau)}$ of the $\alpha_{OX}$ index as a function of the time-lag $\tau$ for the multi-epoch objects in the reference sample. The dotted line is the noise level; the dashed line is the 'uncorrected' structure function (i.e. without noise subtraction); red line is the 'corrected' structure function, where red points are representative values of bins; the blue straight line is a weighted least squares fit to the 'corrected' structure function. Error bars represent the 1-$\sigma$ dispersion of the distribution of points in each bin.}
\label{fig:sf_selected_cut}
\end{figure}

In Figure \ref{fig:sf_selected_cut} it is shown the structure function of the $\alpha_{OX}$ as a function of the time-lag for the objects in the reference sample. The error bars shown in Figure \ref{fig:sf_selected_cut} are not measurement errors but they concern the statistical dispersion of the data in the bins. Indeed, estimating the uncertainties from the observed scatter is the only viable approach for sparsely-sampled light curves characterised by a red-noise behaviour. In fact, as shown by, e.g. \citet{Allevato}, both the photometric errors and the formal uncertainties severely underestimate the scatter intrinsic to any stochastic process.

It can be seen that there is a weak increase of the SF with time-lag. In Figure \ref{fig:sf_selected_cut} it is also shown a weighted least squares fit to the data of the form ${\log{SF(\tau)}=a\log{\tau}+b}$, in which the weight is the number of points in the bin. The result of the fit is $a=0.09{\pm}0.03$ and $b=-1.32{\pm}0.07$. These parameters have been used to estimate the SF value at long time-lags
\begin{equation}
{\log{SF(\tau_{longest})}=a\log{\tau_{longest}}+b}
\label{eqn:weighted_fit}
\end{equation}
where $\tau_{longest}$ is the time-lag value associated to the last bin ($\sim$2000 days).

The SF value at the longest time-lag is $\sim$0.09, and it can be used to constrain the contribution of the intra-source dispersion to the total variance of the relation 
\begin{equation}
{\frac{\sigma^2_{intra-source}}{\sigma^2}{\sim}\left({\frac{0.09}{0.12}}\right)^2{\sim}56\,\%}
\label{eqn:variability}
\end{equation}

This contribution is higher than that found by \citet{Vagnetti2010}. This fact may be due to the longer time-lags sampled in this work ($\sim$2000 days), considering that variability of $\alpha_{OX}$ increases with time-lag, as shown in Figure \ref{fig:sf_selected_cut} and Equation \ref{eqn:weighted_fit}. Indeed, if we evaluate our SF at the time-lag $\sim$300 days as \citet{Vagnetti2010}, the relative contribution of variability is ${\sigma^2_{intra-source}}/{\sigma^2}\sim$44$\%$.

\subsection{The dependence on $L_{UV}$ and $z$}
\label{parametri}

We studied the dependence of the $\alpha_{OX}$ index on the redshift for the reference sample, and we found that the two quantities are negatively correlated: $\alpha_{OX}=(-0.104{\pm}0.008)z+(-0.009{\pm}0.010)$, with a correlation coefficient $r=-0.45$ and a probability $P(>r)\sim8\times\,10^{-3}$ for the null hypothesis that $\alpha_{OX}$ and $z$ are uncorrelated. However, as already suggested by \citet{Vagnetti2010}, this positive correlation between $\alpha_{OX}$ and $z$ may be a by-product of the positive correlation between $L_{UV}$ and $z$. In order to check this possibility, we performed a partial-correlation analysis for the reference sample, and we found a partial correlation coefficient of $\alpha_{OX}$ with the UV luminosity, taking into account the dependence on redshift, $r_{\alpha L,z}=(r_{\alpha L}-r_{\alpha z}r_{zL})/\sqrt{(1-r_{\alpha z}^2)(1-r_{zL}^2)}=-0.59$, with $P(>r)\sim9.5\,10^{-50}$. Similarly, the partial correlation coefficient of $\alpha_{OX}$ with the redshift, accounting for the dependence on the UV luminosity is $r_{\alpha z,L}=(r_{\alpha z}-r_{\alpha L}r_{zL})/\sqrt{(1-r_{\alpha L}^2)(1-r_{zL}^2)}=0.1$, with $P(>r)=0.012$. This result is not as strong as that derived by \citet{Vagnetti2010}, so we can not rule out the possibility of a weak dependence on redshift even taking into account the effect of luminosity. The difference with respect to \citet{Vagnetti2010} may be due to our larger sample. Indeed, referring to figure \ref{fig:pianoLUVz}, we added objects in the low-$z$/low-UV luminosity part of the $z - L_{UV}$ plane, so we may have not added objects uniformly, resulting in a weak redshift dependence. In order to further investigate this possibility, we performed a partial-correlation analysis focusing only on those sources belonging to both \citet{Vagnetti2010} and the reference sample (empty circles with black point inside in Figure \ref{fig:graph_3}), and we found a partial correlation coefficient of $r_{\alpha\,z,L}=-0.09$ with $P(>r)=0.42$, similar to \citet{Vagnetti2010}. This suggests that the result obtained with the reference sample is likely the result of the addition of low-$z$/low-UV luminosity sources.

Then, we divided the sample into two subsamples in redshift and UV luminosity considering the median values $z$=1.28 and $\log{L_{UV}}$=30.26, respectively: they guarantee an approximately equal number of sources in both subsamples. We found $\alpha_{OX}=(-0.214{\pm}0.014)\log{L_{UV}}+(4.96{\pm}0.43)$ for $z>1.28$ sources, with slope in agreement with the high-$z$ sample of \citet{GibsonBrandtSchneider2008}, and $\alpha_{OX}=(-0.150{\pm}0.011)\log{L_{UV}}+(3.01{\pm}0.32)$ for $z<1.28$ sources. Considering the UV luminosities, we found slopes of $(-0.195{\pm}0.014)$ for the $\log{L_{UV}}>30.26$ sample and $(-0.131{\pm}0.014)$ for $\log{L_{UV}}<30.26$, in agreement with \citet{Vagnetti2010}, and similar results has been derived by \citet{Steffen2006}.

\begin{figure}
\includegraphics[width=0.5\textwidth]{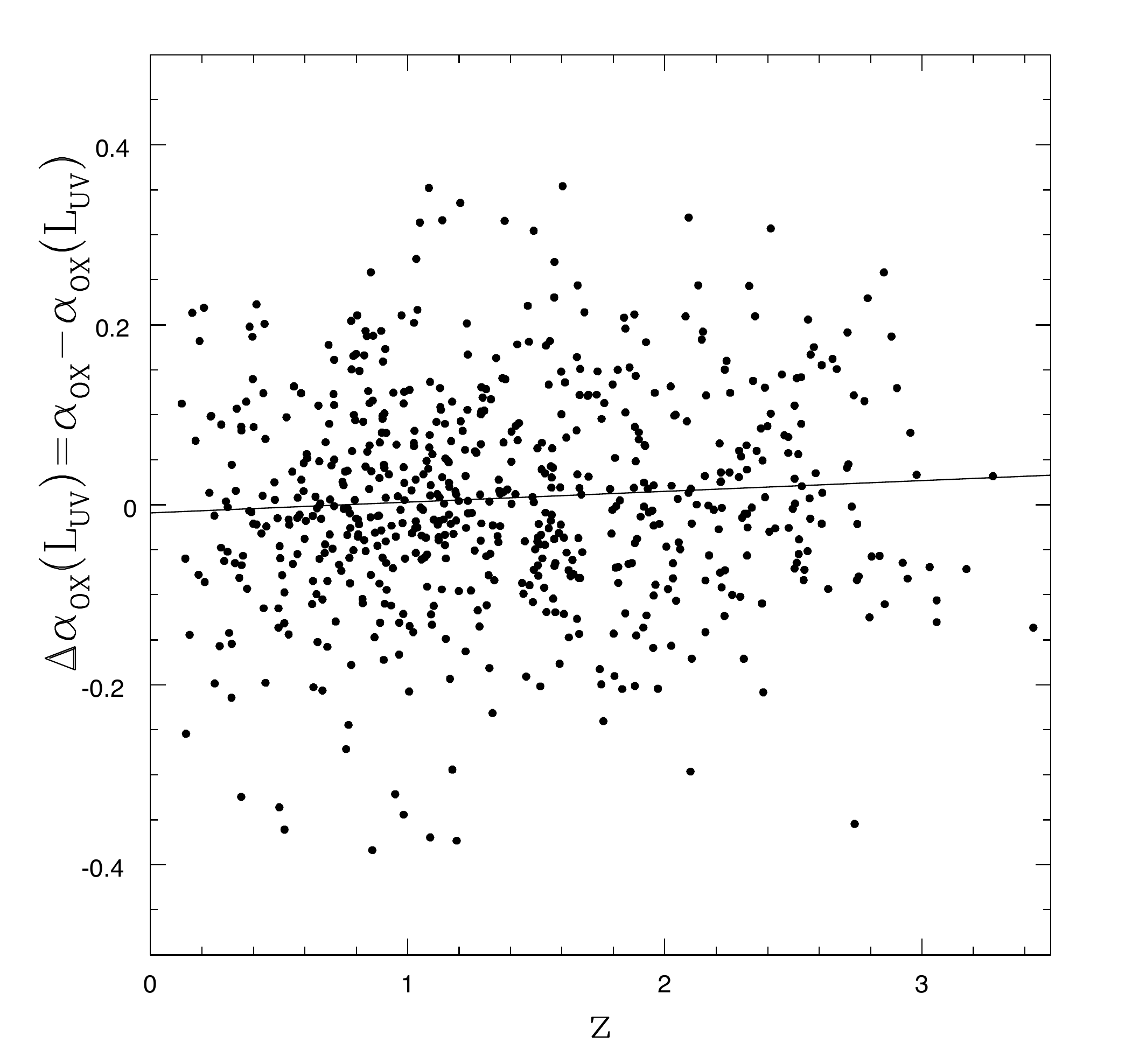}
\caption{Residuals of the $\alpha_{OX} - L_{UV}$ relation as a function of redshift for the objects in the reference sample. The straight line is the linear least squares fit to the data.}
\label{fig:resluv_z}
\end{figure}

We also studied the dependence of the residuals of the  $\alpha_{OX} - L_{UV}$ relation with redshift, and we see (fig. \ref{fig:resluv_z}) that there is a weak and not-significant dependence: 
\begin{equation}
\Delta\alpha_{OX}=(0.011{\pm}0.007)z+(-0.009{\pm}0.010)
\label{eqn:fit_1}
\end{equation}
with a correlation coefficient of ${r=0.07}$ and $P(>r)\sim0.08$. 

Previous works have established that there is essentially no redshift dependence of the relation \citep{Just2007,Vagnetti2010,Vagnetti2013}; however in light of our results we can not rule out a residual dependence on redshift. For the future, larger samples with a wider covering of the $L_{UV} - z$ plane would for sure allow to obtain more robust results.  

Considering the work done by \citet{LussoRisaliti2016} and \citet{LussoRisaliti2017}, we will compare our results obtained when studying the $L_{X} - L_{UV}$ relation with theirs.

\subsection{{The dependence on $M_{BH}$, $L/L_{Edd}$,viewing angle and the origin of inter-source dispersion}}
As already pointed out, our purpose is to deepen the knowledge of the $\alpha_{OX} - L_{UV}$ relation, and most importantly to understand the physical origin of the residual dispersion, i.e. the inter-source dispersion. Indeed, we have found, through variability studies performed via structure function, that an intrinsic variation in the X-ray/UV ratio can account for 56$\%$ of the total variance of the relation. In order to investigate the origin of the residual dispersion, we studied the dependence of $\alpha_{OX}$ and residuals of $\alpha_{OX} - L_{UV}$ relation on fundamental quantities like BH mass, Eddington ratio and we also investigated the role of viewing angle.

\begin{figure}
\includegraphics[width=0.5\textwidth]{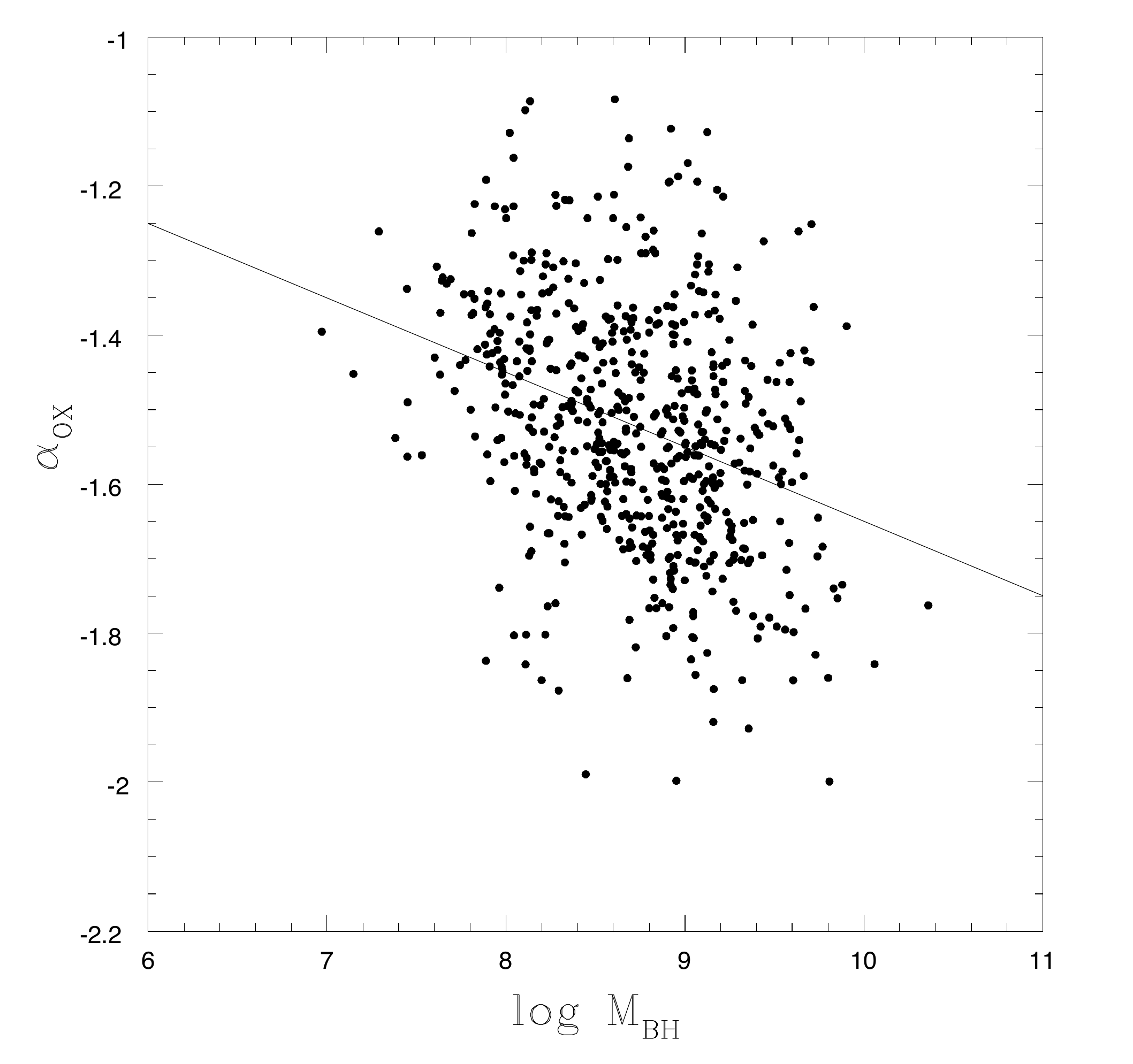}
\caption{$\alpha_{OX}$ index as a function of black-hole mass for the objects in the reference sample. The straight line is the linear least squares fit to the data.}
\label{fig:alphaOX_mbh}
\end{figure}

In Figure \ref{fig:alphaOX_mbh} it is shown the dependence of the $\alpha_{OX}$ index as a function of the BH mass:
\begin{equation}
\alpha_{OX}=(-0.1{\pm}0.01)\log{M_{BH}}+(-0.65{\pm}0.1)
\end{equation}
with a correlation coefficient $r=-0.33$, the probability for the null hypothesis is $P(>r)\sim10^{-17}$. This result is in agreement with \citet{DongGreenHo32012} and can be easily understood considering a standard $\alpha$-disk accretion disk \citep{ShakuraSunyaev1973}: for a fixed bolometric luminosity, a decrease in BH mass results in fainter disk emission in the UV, so higher $\alpha_{OX}$ values.

\begin{figure}
\includegraphics[width=0.5\textwidth]{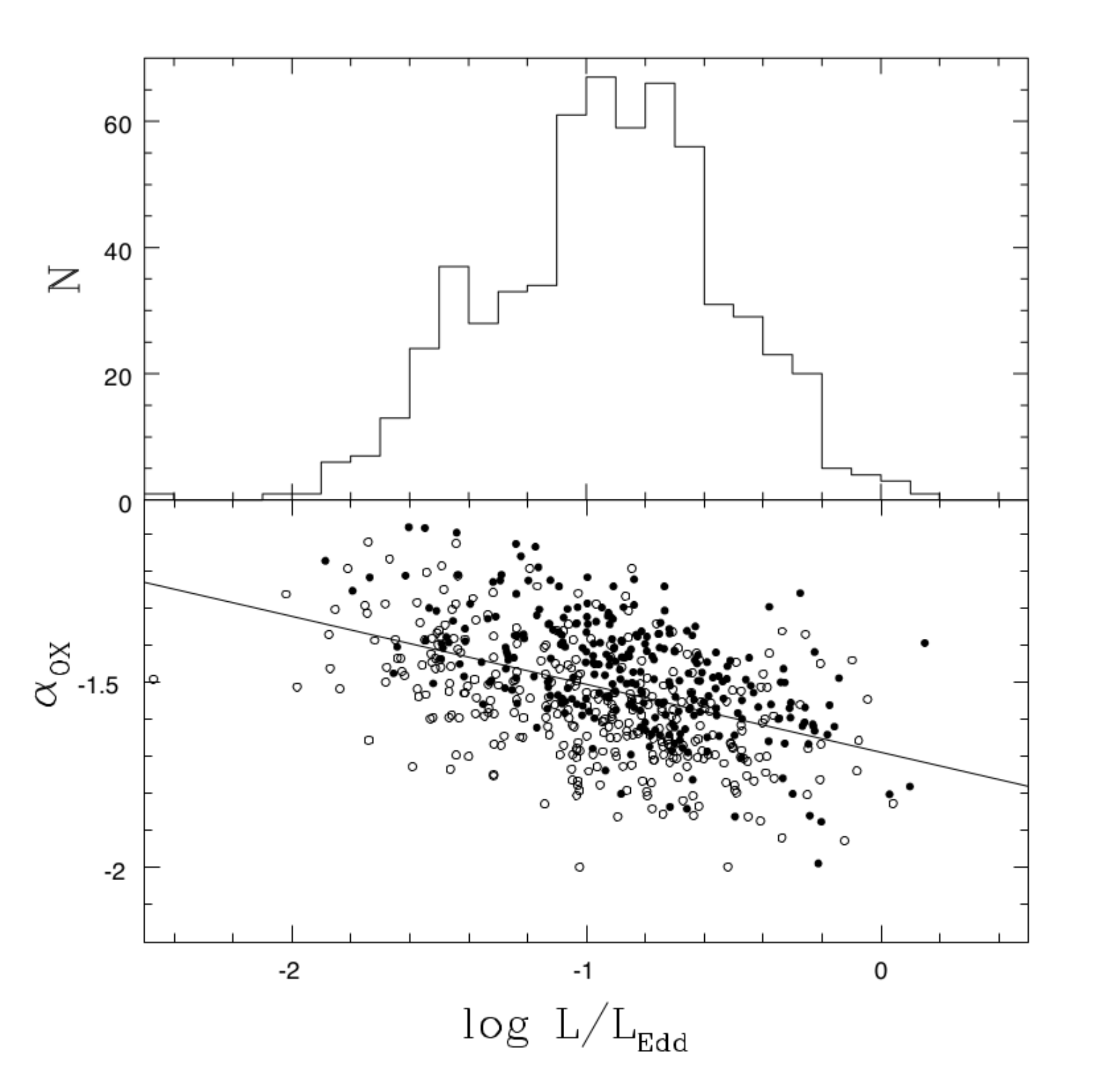}
\caption{Lower Panel: the $\alpha_{OX}$ index as a function of Eddington ratio for the objects in the reference sample. The straight line is the linear least squares fit to the data. Empty circles represent low-$M_{BH}$ objects, filled circles represent high-$M_{BH}$ objects, where the dividing value is the median BH mass of the reference sample $\log{M_{BH}}=8.7$; Top Panel: The distribution of Eddington ratios for the objects in our reference sample.}
\label{fig:alphaOX_etaedd}
\end{figure}

In Figure \ref{fig:alphaOX_etaedd} we show the dependence of $\alpha_{OX}$ as a function of Eddington ratio:
\begin{equation}
\alpha_{OX}=(-0.183{\pm}0.015)\log{L/L_{Edd}}+(-1.69{\pm}0.014)
\end{equation}
with a correlation coefficient $r=-0.45$, $P(>r)\sim10^{-32}$. From Figure \ref{fig:alphaOX_etaedd} we can see that, for a fixed Eddington ratio, objects with higher BH mass have lower X-ray/UV ratios, in agreement with \citet{DongGreenHo32012} and with Figure \ref{fig:alphaOX_mbh}. However, the weak anti-correlation we found between $\alpha_{OX}$ and Eddington ratio is not in agreement with the positive one obtained by \citet{DongGreenHo32012}, and this reflects also in the trend for which, for a fixed BH mass, objects with higher Eddington ratios have on average lower X-ray/UV ratios. This discrepancy may be due to the distribution of Eddington ratios in our sample, with objects being mainly concentrated in a narrow interval $-1\le\,\log{L/L_{Edd}}\le-0.5$ (see Figure \ref{fig:alphaOX_etaedd}). A more homogeneous distribution of Eddington ratios, with more objects populating the high-$L/L_{Edd}$ and low-$L/L_{Edd}$ tails would permit a more robust study on the $\alpha_{OX}$ dependence on this parameter.

The dependence of the $\alpha_{OX}$ on the BH mass could be just a different representation of the dependence on $L_{UV}$, given the correlation between the BH mass and UV luminosity in accretion disk models. Thus, in order to understand the physical origin of the dispersion of the relation, we investigated the dependence of the residuals of the $\alpha_{OX} - L_{UV}$ relation on fundamental quantities. In particular, we studied the dependence of the residuals $\Delta\alpha_{OX}(L_{UV})=\alpha_{OX}-\alpha_{OX}(L_{UV})$ as a function of BH mass and Eddington ratios, and we find weak but significant trends, as follows:
\begin{equation}
{\Delta\alpha_{OX}(L_{UV})=(0.042\pm0.09)\log{M_{BH}}+(-0.36\pm0.08)}
\label{eqn:res_luv_mbh}
\end{equation}

\noindent with r=0.19, P($>$r)$\sim$4$\,$10$^{-6}$, and 
\begin{equation}
{\Delta\alpha_{OX}(L_{UV})=(-0.064\pm0.01)\log{L/L_{Edd}}+(-0.052\pm0.012)}
\label{eqn:res_luv_etaedd}
\end{equation}
with r=-0.21, P($>$r)$\sim\,$10$^{-7}$.

Another way to describe the same dependencies is to consider $\alpha_{OX}$ as dependent on both UV luminosity and the black-hole mass or the Eddington ratio. We have used the macro \textsc{linfit} of the package SM\footnote{\url{https://www.astro.princeton.edu/~rhl/sm/}} which performs a multivariate linear least squares fit, and we have found $\alpha_{OX}=(-0.23\pm0.01)\,\log{L_{UV}}+(0.10\pm0.01){\log{M_{BH}}}+(4.44\pm0.24)$ with $\sigma=$0.11. Considering the Eddington ratio, we have found $\alpha_{OX}=(-0.15\pm0.01)\,\log{L_{UV}}+(-0.07\pm0.01){\log{L/L_{Edd}}}+(2.85\pm0.25)$ with $\sigma=$0.115. As a cross validation, we have performed the same analysis with python package \textsc{scikit-learn} \citep{scikit-learn} and the \textsc{SciPy} \citep{scipy} package \textsc{optimize}, finding consistent values. These results are in agreement with the trends indicated by Equations \ref{eqn:res_luv_mbh} and \ref{eqn:res_luv_etaedd}.

These dependences, although not strong, might be part of the contribution to the inter-source dispersion.

Another possible contribution to the dispersion might come from a spread in corona properties among different sources, as suggested by \citet{DongGreenHo32012}.

At the beginning of this section we suggested that a contribution to the dispersion may be due to the inclination angle, however the problem of finding reliable inclination indicators in AGNs is a hot topic \citep[e.g.][]{Marin2016}. The role of inclination angle has been discussed by \citet{Marziani2018} and \citet{Marziani2001} in light of the Eigenvector 1 (EV1) plane by \citet{BorosonGreen1992}. We refer to figure 2 in \citet{Marziani2018} (but see also figure 1 in \citet{ShenHo2014}) in which it is shown the optical plane of the EV1: $FWHM(H\beta) - R_{Fe_{II}}$, where $R_{Fe_{II}}$ is the ratio of $Fe_{II}$ within $4434{\div}4684{\si\angstrom}$ to broad H$\beta$ EW, $R_{Fe_{II}}=EW(Fe_{II})/EW(H{\beta})$. Following this idea, we made an attempt to built these two quantities for the sample used in this work. Unfortunately, it has been possible to compute the quantities $FWHM(H\beta)$, $EW(Fe_{II})$ and $EW(H\beta)$ for only 50 objects in this sample: they are available only for redshift $z\le0.9$, and are present only in the catalogue by \citet{Shen2011}, not in the one by \citet{Kozlowski2017}. However, according to \citet{ShenHo2014}, the dispersion in the FWHM(H$\beta$) is mainly attributed to an inclination effect, which makes this parameter a reliable inclination indicator. Thus, we correlated the $\alpha_{OX}$ with the $FWHM(H{\beta})$ for the 54 objects in the Reference Sample which were provided with estimates of the $FWHM(H{\beta})$. We obtained:
\begin{equation}
{\alpha_{OX}=(0.16{\pm}0.08)\log{FWHM(H{\beta})}+(-2.01{\pm}0.29)}
\end{equation}
with a correlation coefficient of ${r=0.26}$ and ${P(>r)=0.06}$.

This positive correlation shown is in agreement with scenario depicted by \citet{You2012}. They built up a general-relativistic (GR) model for an accretion disk + corona model sorrounding a Kerr black-hole, in which the inclination angle plays a crucial role: the emission from the corona can be approximated to be isotropic while the emission from the accretion disk is directional, resulting in an increase of the X-ray/UV ratio with viewing angle. 

\begin{figure}
\includegraphics[width=0.5\textwidth]{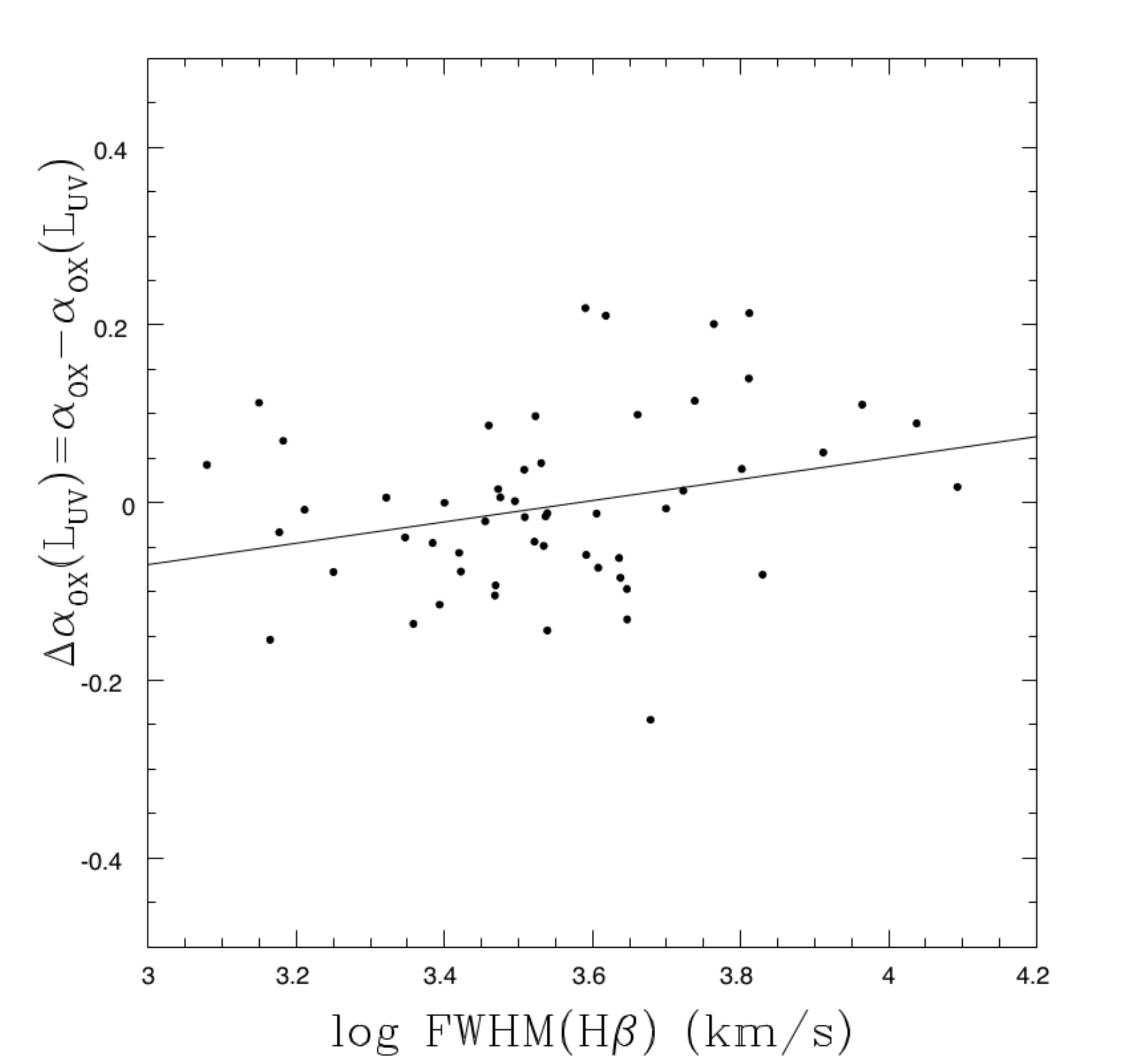}
\caption{Residuals of the $\alpha_{OX} - L_{UV}$ relation as a function of the $FWHM(H\beta)$ for the objects belonging to the reference sample and for which estimates of the $FWHM(H\beta)$ are available.}
\label{fig:res_alphaOX_fwhm_hb}
\end{figure}

However, our purpose is to investigate contribution of fundamental physical quantities to the dispersion of the $\alpha_{OX} - L_{UV}$ relation. For this reason, we have studied the dependence of the residuals of the above relation as a function of the $FWHM(H\beta)$ (see Figure \ref{fig:res_alphaOX_fwhm_hb}):
\begin{equation}
{\Delta{\alpha_{OX}}=(0.12{\pm}0.06)\log{FWHM(H{\beta})}+(-0.43{\pm}0.21)}
\end{equation}
with a correlation coefficient of $r=0.27$ and $P(>r)=0.048$.

We note that the EW[O$_{III}$] has been also identified as an orientation indicator \citep{Risaliti2011,Bisogni2017}. We have therefore correlated our data with this parameter for the Reference Sample, finding significant correlations as follows:
\begin{equation}
{\alpha_{OX}=(0.17\pm0.04)\log{EW[O_{III}]}+(-1.65\pm0.05)}
\end{equation}
\noindent with r=0.5 and P($>$r)$\sim2\,$10$^{-4}$, and 
\begin{equation}
{\Delta\alpha_{OX}(L_{UV})=(0.10\pm0.03)\log{EW[O_{III}]}+(0.12\pm0.04)}
\end{equation}
\noindent with r=0.4 and P($>$r)$\sim5\,$10$^{-3}$, in agreement with the trend found for the case of the FWHM(H$\beta$).

Our results, although not statistically robust, indicate a possible interesting trend, and for the future, a sample for which estimates of the three quantities are available for a larger number of objects might allow a quantitative study of the impact of inclination to the dispersion of the $\alpha_{OX} - L_{UV}$ relation. Indeed, such a sample might allow a division of the sample with respect to viewing angle and the selection of sources expected to contribute less to the dispersion of the relation based on their inclination angle.

\section{The $L_{X} - L_{UV}$ relation and its use in cosmology}
We have mentioned before that the $\alpha_{OX} - L_{UV}$ relation is a byproduct of the well-established positive correlation between $L_{X}$ and $L_{UV}$ luminosity. This relation has been studied thoroughly by Lusso and Risaliti in a series of papers as they use it to build a Hubble diagram for quasars \citep{Risaliti2015,LussoRisaliti2016,LussoRisaliti2017,BisogniRisalitiLusso2017}. Indeed, \citet{Risaliti2015} considered objects from SDSS cross-matched with samples with X-ray measurements from literature and used the non-linear relation between X-ray and UV luminosity to build a Distance Modulus (DM) - redshift plane and estimate cosmological parameters $\Omega_m$ and $\Omega_{\Lambda}$. In a more recent paper, \citet{BisogniRisalitiLusso2017} built the same diagram for a sample of $8000$ objects obtained by cross-matching catalogues by \citet{Shen2011} and \citet{Paris2017} with 3XMM-DR5 \citep{Rosen2016}, and in both works they found values for the parameters in agreement with those derived from the well-known Hubble diagram built from supernovae.

This method represents a valid alternative to the supernovae diagram, and it has several advantages with respect to it: it can be used at higher redshifts (up to $\sim5$), and it has a better statistics. However, the use of the $L_{X} - L_{UV}$ relation to build a Hubble diagram for quasars relies on the tightness of the relation, and so the DM - $z$ diagram created will be charachterised by a larger dispersion with respect to supernovae. Nevertheless, \citet{LussoRisaliti2016} proved that by carefully selecting the sample it is possible to decrease the dispersion of the relation and so legitimate the use of this relation for cosmological purposes. In our work, we applied constraints on our initial sample in order to decrease as much as possible the dispersion of the relation and we further analysed the data to understand the physical origin of the residual dispersion of the relation. Indeed, it is clear that a thorough study on the dispersion and its origin is of vital importance for the use of the relation in cosmology in the sense described above.

With the reference sample described above we studied the relation between X-ray and UV luminosity. However, while in the case of the $\alpha_{OX} - L_{UV}$ relation the $\alpha_{OX}$ index is assumed the dependent variable, the $L_{UV}$ being the independent one, a linear least squares fit turns out to be a good fit. When considering the $L_X - L_{UV}$ relation it is not possible to consider $L_{X}$ the dependent variable and $L_{UV}$ the independent one (or viceversa), because we still lack a full understanding of the relation between the two regions responsible for the emission and because the relation is affected by large dispersion, so other methods must be employed \citep{LussoRisaliti2016,Tang2007}. We used the Orthogonal Distance Regression (ODR) fitting\footnote{\url{http://docs.scipy.org/doc/scipy/reference/odr.html}}. The ODR fitting method treats X and Y variables symmetrically and minimises both the sum of the squares of the X and Y residuals, taking into account uncertainties in both variables. The result of the ODR fitting to the data is
\begin{equation}
\log{L_X}=(0.671{\pm}0.013)\log{L_{UV}}+(6.15{\pm}0.40)
\label{eqn:lxluv}
\end{equation}
Our slope is comparable with that derived by \citet{LussoRisaliti2016}, $0.634{\pm}0.013$, although our dispersion of $\sigma\sim0.31$ is higher than that derived by those authors, due to some differences in the analyses.
Indeed, \citet{LussoRisaliti2016} use non-simultaneous UV and X-ray measurements. On one side,  this has the advantage of better photometry, both in the X-rays (using the longest exposures) and in the UV  (using the SDSS photometry which takes into account emission lines). On the other side, our simultaneous analysis is capable of a better treatment of variability via the appropriate use of the SF. In fact, we find a larger contribution of variability, which translates into a residual dispersion (inter-source dispersion) which is similar to that of \citet{LussoRisaliti2016}.

We notice that the dispersion can be further reduced by selecting only those sources having a large number of observational epochs, as discussed at the end of Section 3.1. For the future, a more precise determination of the Equations \ref{eqn:fit} and \ref{eqn:lxluv} might come adopting samples containing only objects with a large number of epochs.

\section{Discussion and conclusions}

The purpose of our work is to estimate the contribution of intrinsic variation of the X-ray/UV ratio to the dispersion of the $\alpha_{OX} - L_{UV}$ relation, and in particular to understand the origin of the residual dispersion of the relation with simultaneous X-ray and UV observations coming from the MEXSAS2 catalogue and the XMM-SUSS3, respectively. Indeed, once simultaneous X-ray and UV observations are used, the dispersion of the $\alpha_{OX} - L_{UV}$ relation is given by two contributions: an intra-source dispersion, due to intrinsic variations in the X-ray/UV ratio in single sources, and an inter-source dispersion, which may be due to fundamental quantities like BH mass, Eddington ratio, and/or viewing angle.

Starting from the parent sample, which is the result of the cross-match between MEXSAS2 and the XMM-SUSS3, we applied stringent constraints in order to decrease as much as possible the dispersion of the relation, following the strategy adopted by \citet{LussoRisaliti2016}. We considered only non-BAL and non-RL objects, as they would increase the dispersion, we took into account for the effects of intergalactic $H_{I}$ absorption and extinction, and we considered only non-absorbed (in X-rays) sources with reliable photon-index estimates.  

We have shown that by carefully selecting the sample with the constraints described above, it is possible to decrease the dispersion of the $\alpha_{OX} - L_{UV}$ relation, in agreement with \citet{LussoRisaliti2016}. We have confirmed the negative correlation between the two quantities, with a slope of  -0.159${\pm}$0.007, comparable to slopes obtained by other autors \citep[e.g.][]{Just2007,Lusso2010,Vagnetti2010}, and we obtained a dispersion of $\sim$0.12, consistent with \citet{Vagnetti2010}. 

Moreover, we performed an ensemble variability analysis of the $\alpha_{OX}$ index by means of the Structure Function. Indeed, the variance of the $\alpha_{OX} - L_{UV}$ relation can be written as the sum of two contributions, an \textit{intra-source} and an \textit{inter-source} dispersion, and from the SF value at long time-lags we estimated that a true variability in the X-ray/UV ratio contributes for the 56$\%$ to the total variance of the relation (intra-source dispersion).

Considering \citet{LussoRisaliti2016}, they found for the $L_{X} - L_{UV}$ relation a residual dispersion of  $\sigma\sim0.19$, i.e. dispersion which is not explained by a true variability in the X-ray/UV ratio. Our result means that the dispersion which cannot be explained with a true variability in the X-ray/UV ratio is approximately ${\sim{\sqrt{1-0.56}}\sigma\sim0.2}$ (see Equation \ref{eqn:variability}), similar to that derived by \citet{LussoRisaliti2016}.

In order to decrease the dispersion of the relation, we made an attempt by removing sources with only one or two observations, finding that it can decrease by approximately 15\%.

The residual dispersion in the relation may be due to other physical quantities, like black-hole mass, Eddington ratio and inclination angle.

First, we studied the dependence of the relation on redshift and optical/UV luminosity. We have performed a partial correlation analysis for the $\alpha_{OX} - L_{UV}$ relation taking into account the effect of redshift and for the $\alpha_{OX} - z$ relation taking into account the effect of UV luminosity, and we found $r_{\alpha\,z,L}=0.1$ with $P(>r)=0.012$: our result is not as strong as previous works \citep[e.g.][]{Just2007,Vagnetti2010,Vagnetti2013}, so we can not rule out a residual dependence on redshift. For the future, larger samples with wider and more uniform covering of the $L_{UV} - z$ plane will allow to obtain more robust results in this sense.

Second, we studied the dependence of the residuals of the $\alpha_{OX} - L_{UV}$ relation on black-hole mass and Eddington ratio, and of the $\alpha_{OX}$ index on these quantities. We have found weak but significant trends indicating an increase of the residuals of the relation with black-hole mass and a decrease with Eddington ratio. However, the dependence on these quantities may be masked by the dependence on UV luminosity. To test this issue, we performed a multivariate regression analysis considering $\alpha_{OX}$ as a function of UV luminosity and black-hole mass or Eddington ratio. The results we have found are in agreement with the trends in the residuals.

We also studied the dependence of the $\alpha_{OX}$ index and the residuals of the $\alpha_{OX} - L_{UV}$ relation on the inclination angle, and we considered as indicator the FWHM(H$\beta$), following \citet{Marziani2001,Marziani2018} and \citet{ShenHo2014}. We have found that the residuals of the relation and the $\alpha_{OX}$ index are positively correlated with  FWHM(H$\beta$), with slopes of 0.13${\pm}$0.06 and 0.18$\pm$0.09, respectively, with the latter result in agreement with the scenario depicted by \citet{You2012}, according to which, in a GR model of an accretion disk+corona around a Kerr black-hole, higher inclination-angle objects would be characterised by higher $\alpha_{OX}$ values. We have performed the same analysis considering another inclination indicator, the EW[O$_{III}$], and we have found similar results. However, due to the poor statistics of our sample when considering the two quantities, these results are not robust. Nevertheless, they can represent a starting point for possible future studies. Indeed, a sample for which estimates of the FWHM(H$\beta$) (as well as EW[O$_{III}$]) are available for a larger number of objects, uniformly distributed in inclination angle, would for sure allow more robust studies. In particular, in light of the use of the $L_{X} - L_{UV}$ relation in cosmology, it would allow the possibility to divide the sample in intervals of inclination and select only those objects characterised by low values of the residuals, in order to decrease the dispersion of the relation.

\begin{acknowledgements}
{We are grateful to the referee whose comments improved the quality of this work}. 
E.C. acknowledges the National Institute of Astrophysics (INAF) and the University of Rome - Tor Vergata for the PhD scholarships in the XXXIII PhD cycle.
F.T. acknowledges support by the Programma per Giovani Ricercatori - anno 2014 “Rita Levi Montalcini”.
F.V. and M.P. acknowledge support from the project \textit{Quasars at high redshift: physics and Cosmology} financed by the ASI/INAF agreement 2017-14-H.0.
This research has made use of data obtained from the 3XMM-Newton serendipitous source catalogue compiled by the 10 institutes of the XMM-Newton Survey Science Centre selected by ESA.
This research has made use of the XMM-OM Serendipitous Ultra-violet Source Survey, which has been created at the University College London’s (UCL’s) Mullard Space Science Laboratory (MSSL) on behalf of ESA and is a partner resource to the 3XMM serendipitous X-ray source catalog.

\end{acknowledgements}

\bibliographystyle{aa}
\bibliography{biblio}

\end{document}